\documentclass[conference]{IEEEtran}
\IEEEoverridecommandlockouts
\usepackage{cite}
\usepackage{amsmath,amssymb,amsfonts}
\usepackage{graphicx,pdfpages,xcolor}
\usepackage{braket}
\usepackage{bm}
\usepackage{balance}

\def\BibTeX{{\rm B\kern-.05em{\sc i\kern-.025em b}\kern-.08em
    T\kern-.1667em\lower.7ex\hbox{E}\kern-.125emX}}

\newtheorem{theorem}{Theorem}
\newtheorem{lemma}{Lemma}

\newtheorem{corollary}{Corollary}

\newcommand{\Var}[1]{\textnormal{\textsf{Var}}\!\left({#1}\right)} %variance

\newcommand{\Exp}{\textnormal{\textsf{E}}} %expectation (no argument)
\newcommand{\E}[2][]{\textnormal{\textsf{E}}_{#1}\!\left[#2\right]} %expectation
\newcommand{\Prob}{\textnormal{Pr}} 
\DeclareMathOperator{\Tr}{Tr}

\title{A Converse Bound via the Nussbaum–Szkoła Mapping
for Quantum Hypothesis Testing
\thanks{The authors have received funding from the Spanish Ministerio de Ciencia, Innovación y Universidades under Grant PID2024-159557OB-C21 (MICIU/AEI/10.13039/501100011033 and ERDF/UE) and from the Comunidad de Madrid under Grant IDEA-CM (TEC-2024/COM-89).}}
\author{}
\date{}

\author{%
  \IEEEauthorblockN{Jorge Lizarribar-Carrillo\IEEEauthorrefmark{1}\IEEEauthorrefmark{2}, Gonzalo Vazquez-Vilar\IEEEauthorrefmark{1}\IEEEauthorrefmark{2}, and Tobias Koch\IEEEauthorrefmark{1}\IEEEauthorrefmark{2}}
     \IEEEauthorblockA{\IEEEauthorrefmark{1}%
             Signal Theory and Communications Department, Universidad Carlos III de Madrid, 28911 Legan\'es, Spain}
  \IEEEauthorblockA{\IEEEauthorrefmark{2}%
             Gregorio Mara\~n\'on Health Research Institute, 28007 Madrid, Spain}
             \IEEEauthorblockA{Emails: jlizarri@ing.uc3m.es, gonzalo.vazquez@uc3m.es, tkoch@ing.uc3m.es}
}%\author

\begin{document}

\maketitle

\begin{abstract}
Quantum hypothesis testing concerns the discrimination between quantum states. This paper introduces a novel lower bound for asymmetric quantum hypothesis testing that is based on the Nussbaum–Szkoła mapping. The lower bound provides a unified recovery of converse results across all major asymptotic regimes, including large-, moderate-, and small-deviations. Unlike existing bounds, which either rely on technically involved information-spectrum arguments or suffer from fixed prefactors and limited applicability in the non-asymptotic regime, the proposed bound arises from a single expression and enables, in some cases, the direct use of classical results. It is further demonstrated that the proposed bound provides accurate approximations to the optimal quantum error trade-off function at small blocklengths. Numerical comparisons with existing bounds, including those based on fidelity and information spectrum methods, highlight its improved tightness.
\end{abstract}

%\begin{IEEEkeywords}
%Quantum hypothesis testing; finite blocklength; small deviations analysis; Nussbaum–Szkoła mapping.
%\end{IEEEkeywords}

\section{Introduction}

Hypothesis testing is a central problem in information theory, concerned with distinguishing between two competing sources based on observed data. In the quantum setting, this task amounts to distinguishing between copies of two quantum states using measurements described by positive operator-valued measures (POVMs). The non-commutativity of quantum states gives rise to phenomena with no classical analogue, making the characterization of optimal error probabilities substantially more challenging.

A binary hypothesis test involves two types of errors: a \mbox{type-I} error, which occurs when the null hypothesis is incorrectly rejected, and a type-II error, which occurs when it is incorrectly accepted. The fundamental trade-off between these errors is typically formalized by optimizing one error probability subject to a constraint on the other.
In the quantum setting, this trade-off admits an exact characterization via a semidefinite program (SDP). 
However, the dimension of the resulting SDP grows rapidly with both the dimension of the quantum states and the number of copies.
This rapid scaling renders exact computations infeasible beyond very small systems, thereby motivating the development of analytically tractable and computationally efficient bounds that capture the essential behavior of the error probability.

Two lower bounds are particularly relevant in this context: 
The first, established by Nussbaum and Szkoła~\cite{nussbaum2009lower}, provides a fundamental link between quantum and classical hypothesis testing. This result bounds the optimal average error probability by one-half of the classical optimal error probability associated with the so-called Nussbaum–Szkoła distributions, which  are explicitly constructed from the spectral decompositions of the original quantum states $\rho$ and $\sigma$, effectively embedding the non-commutative geometry of the Hilbert space into a joint classical distribution. 
The second, proposed by Pereira \emph{et al.}~\cite{pereira2023analytical}, provides a lower bound in the asymmetric case based on the quantum fidelity $0 \leq F(\rho,\sigma) \leq 1$, which quantifies the overlap between the two quantum states.

For large numbers of copies~$n$, asymptotic methods characterize the type-II error probability $\beta_n$ under a prescribed scaling of the type-I error $\alpha_n$. The \textit{small-deviations regime} covers the asymptotic analysis of $\beta_n$ when $\alpha_n$ is fixed. The quantum Stein's lemma establishes the error exponent in this setting \cite{hiaipetz1991, ogawahayashi2004}, thereby providing the leading term of order $n$ in the expansion of $-\log \beta_n$. A second-order analysis refines this result by characterizing the correction term of order $\sqrt{n}$. In the quantum setting, two pioneering works approach this problem from different perspectives: Li~\cite{li2014second} employs elementary algebraic techniques and probability bounds derived from the Nussbaum–Szkoła mapping to characterize the second-order term, while Tomamichel and Hayashi~\cite{tomamichel2013hierarchy} use one-shot entropic quantities---such as the information spectrum relative entropy---to obtain the same result.

When the type-I error decays subexponentially, the problem falls in the \textit{moderate-deviations regime}. The exponent of the type-II error was analyzed asymptotically in~\cite{chubb2017moderate, cheng2017moderate}. In \cite{chubb2017moderate}, Chubb \emph{et al.} used a combination of relative entropy inequalities~\cite{tomamichel2013hierarchy} and a Berry–Esseen–type result due to Rozovsky~\cite{rozovsky2002estimate}. In~\cite{cheng2017moderate}, Cheng and Hsieh derived the achievability by using a martingale inequality due to Sason~\cite{sason2012moderate}, originally developed for classical moderate deviations, while their converse follows directly from a sharp converse Hoeffding bound.

When both errors decay exponentially, the problem falls in the \emph{large-deviations regime}, with performance described by the error exponents $A = -\lim_{n\to\infty}\tfrac{1}{n}\log \alpha_n$ and $B = -\lim_{n\to\infty}\tfrac{1}{n}\log \beta_n$. In the quantum setting, the converse part of this analysis was established by Nussbaum and Szkoła~\cite{nussbaum2009lower}, while the achievability was proven by Audenaert \emph{et al.}~\cite{audenaert2007discriminating}, together yielding the quantum generalization of the classical Chernoff exponent. The Hoeffding bound in the asymmetric setting was established by Hayashi and Nagaoka~\cite{hayashi2006error,nagaoka2006converse}.

%\subsection{Contributions}
%\newpage
In this work, we propose a novel converse bound for asymmetric quantum hypothesis testing based on the Nussbaum–Szkoła mapping. The bound is parametrized by a scalar $0 \leq s < 1$ and provides a unified and conceptually simple framework that recovers converse results across the main asymptotic regimes: large-, moderate-, and small-deviations.
This unification follows from a single expression by selecting appropriate sequences of the parameter $s \equiv s_n$. In certain regimes, it further enables a direct transfer of classical hypothesis testing results to the quantum setting.
We further show that our bound is asymptotically tight for pure, asymptotically orthogonal states, implying that the associated multiplicative constants cannot be improved in general.

%The proposed bound improves upon several previously known results. While the bounds of Nussbaum and Szkoła \cite{nussbaum2009lower} and their asymmetric extension \cite{vazquez2024error} correctly recover the converse exponents in the Hoeffding, Chernoff, and moderate-deviation regimes, their fixed prefactors prevent them from capturing the correct behavior in the small-deviation regime. Information-spectrum-based bounds due to Tomamichel and Hayashi \cite{tomamichel2013hierarchy} fully characterize the moderate- and small-deviation regimes, but rely on technically involved derivations and exhibit limited accuracy at small blocklengths. Fidelity-based bounds derived from Fuchs–van de Graaf inequalities \cite{pereira2023analytical}, while tight for pure states and computationally efficient, are inherently limited in range by the value of the fidelity.
%In contrast, the lower bound introduced here simultaneously recovers the correct converse behavior in the different asymptotic regimes, provides accurate approximations to the quantum error trade-off function at small blocklengths, and outperforms existing efficiently computable bounds in a broad range of scenarios.

%\section{Main Result}

\section{Preliminaries}

We study the problem of discriminating between two quantum states. Specifically, let us consider the density operators\footnote{Density operators are self-adjoint, positive semidefinite, and have unit trace.} $\rho$ and $\sigma$,  acting on some finite-dimensional complex Hilbert space~$\mathcal{H}$ with dimension $d$, and define the hypotheses 
\begin{align}
 \mathrm{H}_0\colon \rho,\qquad
 \mathrm{H}_1\colon \sigma.
\end{align}
In this binary setting, we distinguish between two error types:
\begin{itemize}
    \item The type-I error is the error of accepting $\mathrm{H}_1$ when the true state is the null hypothesis $\mathrm{H}_0\colon \rho$.
    \item The type-II error is the error of accepting $\mathrm{H}_0$ when the true system state is the alternative hypothesis $\mathrm{H}_1\colon \sigma$.
\end{itemize}

A binary test is defined by a positive self-adjoint operator $\Pi$ acting on~$\mathcal{H}$ such that $0\preceq \Pi \preceq I$, where $I$ denotes the identity matrix and the notation $A \preceq B$ means that $B-A$ is positive semidefinite. For a test $\Pi$ associated with $\mathrm{H}_1$, let $\bar\Pi \triangleq I-\Pi$. The type-I and type-II error probabilities are, respectively,
\begin{align}
  \alpha(\Pi) = \Tr[\Pi \rho], \qquad \beta(\Pi) = \Tr[\bar\Pi \sigma] = 1 - \Tr[\Pi \sigma].
\end{align}
The two error probabilities cannot be made arbitrarily small at the same time.
The best achievable trade-off between these probabilities corresponds to the Pareto-optimal boundary
\begin{align}
  \beta_{\alpha}(\rho,\sigma) = \inf_{\Pi: \alpha(\Pi) \leq \alpha} \beta(\Pi).
  \label{eqn:beta-alpha-def}
\end{align}
For classical distributions $P$ and $Q$, $\beta_{\alpha}(P,Q)$ is defined analogously.
The error trade-off admits the following formulation:
\begin{lemma}[\hspace{-0.01mm}{\cite[Lemma~2]{vazquez2016multiple}}]\label{lemma:1}
Let  $\rho$ and $\sigma$ be two density operators. Then,
\begin{align}
\beta_{\alpha}(\rho, \: \sigma) = \sup_{t\geq 0} \left\{\Tr[\sigma \Pi_t] + t \left( \Tr[\rho\bar{\Pi}_t] - \alpha\right) \right\}, \label{eq:var}
\end{align}
where $\Pi_t \triangleq \{ t\rho - \sigma \succeq 0\}$ is the projection onto the non-negative eigenspace of $t\rho - \sigma$ and $\bar{\Pi}_t = I - \Pi_t$.
\end{lemma}

When the quantum states are $n$-fold tensor products, i.e., $\rho \equiv \rho^{\otimes n}$ and $\sigma \equiv \sigma^{\otimes n}$, we consider the asymptotic behaviors of the type-I and type-II error probabilities as $n\to\infty$.
%\footnote{Here $\rho^{\otimes n}$ denotes the $n$-fold tensor product of some basis state $\rho$ and it corresponds to a density operator itself.}
%Several of these asymptotic results were obtained using a mapping, first proposed by Nussbaum and Szkoła in~\cite{nussbaum2006lower}, that relates the quantum testing problem to a classical one with the same asymptotic exponential behavior.

% \vspace{0.5 cm}

%In this work, we study the Nussbaum-Szkoła mapping in the non-asymptotic setting of fixed $n$. We analyze its properties and highlight the distinctions between quantum and classical testing problems through specific examples.

\section{Error Probability Lower Bound}

\subsection{The Lower Bound}

The Nussbaum-Szkoła mapping relates two quantum states $\rho$ and $\sigma$ to two classical distributions $P$ and $Q$. Specifically, for the spectral decompositions of the quantum states $\rho$ and $\sigma$
 \begin{align}
\rho = \sum_{i=1}^d \lambda_i \ket{x_i} \bra{x_i}, \quad \sigma = \sum_{j=1}^d \mu_j\ket{y_j}\bra{y_j} \label{eq:spec}
\end{align}
the Nussbaum-Szkoła-mapping distributions are given by
\begin{align}
P_{ij} \triangleq \lambda_i \lvert\braket{x_i | y_j}\rvert^2, \quad Q_{ij} \triangleq \mu_j\lvert\braket{x_i|y_j}\rvert^2. \label{eq:Ns-mapping}
\end{align}
The next theorem relates the quantum and classical hypothesis testing trade-off functions of \eqref{eq:spec} and \eqref{eq:Ns-mapping}, respectively.
Specifically, the Pareto-optimal boundary of the quantum test  \eqref{eq:spec} is lower bounded by the Pareto-optimal boundary of the test \eqref{eq:Ns-mapping}, with the type-I and type-II error probabilities weighted by the respective factors $1-s$ and $s$, for an arbitrary $0 \leq s < 1$.

\begin{theorem}\label{thm:low}
For a binary quantum hypothesis test between states $\rho$ and $\sigma$ \eqref{eq:spec} with Nussbaum-Szkoła mapping distributions $P$ and $Q$ \eqref{eq:Ns-mapping}, we have that
\begin{align}
\beta_{\alpha}(\rho, \sigma)  &\geq s \beta_{\frac{1}{1-s}\alpha}(P, Q), \label{eq:thmlow}
\end{align}
for every $0 \leq s < 1$ and $0 \leq \alpha \leq 1-s$.
\end{theorem}
%The lower bound in Theorem~\ref{thm:low} admits a natural operational interpretation in terms of the optimal trade-off achieved by the corresponding classical hypothesis test between the distributions $P$ and $Q$. Specifically, the lower bound corresponds to the optimal tradeoff of the classical test with the type-I and type-II error probabilities are weighted by the factors $1-s$ and $s$, respectively.
\begin{IEEEproof} Using  that $\sum_{i=1}^d\!\ket{x_i}\bra{x_i}$ and $\sum_{j=1}^{d}\ket{y_j}\bra{y_j}$ are equal to the identity matrix, together with the idempotency of $\Pi_t$ and $\bar{\Pi}_t$ alongside the cyclic property of the trace, we rewrite the trace terms in Lemma~\ref{lemma:1} as
\begin{align}
    \Tr[\sigma \Pi_t] &= \sum_{ij} \mu_j \lvert \braket{x_i | \Pi_t | y_j}\rvert^2, \label{eq:term1} \\
    \Tr[\rho \bar{\Pi}_t] &= \sum_{ij} \lambda_i \lvert \braket{x_i | \bar{\Pi}_t | y_j}\rvert^2. \label{eq:term2} 
\end{align} 
Substituting the identities \eqref{eq:term1} and \eqref{eq:term2} into the variational expression of Lemma~\ref{lemma:1}, we obtain
\begin{align}
\beta_\alpha(\rho, \: \sigma) \geq \sum_{ij} \mu_j \lvert \braket{x_i | \Pi_t | y_j}\rvert^2 + \sum_{ij} t\lambda_i \lvert \braket{x_i | \bar{\Pi}_t | y_j}\rvert^2 - t\alpha
\end{align}
for any fixed $t\geq 0$. Let $t = \delta t'$ for $\delta \geq 0$. After joining both sums, and lower-bounding $t'\lambda_i$ and $\mu_j$ by $\min(t'\lambda_i, \mu_j)$, this can be further lower-bounded by
\begin{align}
\beta_\alpha(\rho, \: \sigma) \geq \sum_{ij} \min(t'\lambda_i, \mu_j)\Bigl( \lvert \braket{x_i | \Pi_t | y_j}\rvert^2  \bigr. \notag \\ {} + \bigl. \delta \lvert \braket{x_i | \bar{\Pi}_t | y_j}\rvert^2 \Bigr) - \delta t'\alpha. \label{eq:ineqdelta}
\end{align}

Next, for an arbitrary $0\leq s < 1$, define the vectors
\begin{align}
u & \triangleq \left[\sqrt{s}, \: \sqrt{1-s} \right],\\
v & \triangleq \left[\lvert\braket{x_i | \Pi_t | y_j}\rvert, \: \sqrt{\delta} \lvert \braket{x_i | \bar{\Pi}_t | y_j}\rvert \right].\label{eq:vecs}
\end{align}
The Cauchy-Schwarz inequality $\lvert\braket{u,v}\rvert^2 \leq \braket{u,u} \braket{v,v}$ yields
\begin{align}
    \left(\sqrt{s}\lvert\braket{x_i | \Pi_t | y_j}\rvert +  \sqrt{1-s}\sqrt{\delta} \lvert \braket{x_i | \bar{\Pi}_t | y_j}\rvert \right)^2 \notag \\
    \leq \lvert\braket{x_i | \Pi_t | y_j}\rvert^2 +  \delta \lvert \braket{x_i | \bar{\Pi}_t | y_j}\rvert^2. \label{eq:cauchyschwarz}
\end{align}
By choosing $\delta = s/(1 - s)$ in \eqref{eq:cauchyschwarz}, we obtain
\begin{align}
    &s \left(\lvert\braket{x_i | \Pi_t | y_j}\rvert + \lvert \braket{x_i | \bar{\Pi}_t | y_j}\rvert \right)^2 \notag \\ &\qquad\leq\lvert\braket{x_i | \Pi_t | y_j}\rvert^2 + \frac{s}{1-s} \lvert \braket{x_i | \bar{\Pi}_t | y_j}\rvert^2. \label{eq:cauchyschwarz2}
\end{align}
This particular choice of $\delta$ enforces $\delta \geq 0$ for $0 \leq s < 1$. We next note that, by the triangle inequality $\lvert u_1\rvert + \lvert u_2\rvert \geq \lvert u_1 + u_2\rvert$ and the fact that $\Pi_t + \bar{\Pi}_t = I$,
\begin{align}
\lvert\braket{x_i | \Pi_t | y_j}\rvert + \lvert \braket{x_i | \bar{\Pi}_t | y_j}\rvert \geq \lvert\braket{x_i | y_j}\rvert. \label{eq:Tobis_5cents}
\end{align}
Applying \eqref{eq:cauchyschwarz2} and \eqref{eq:Tobis_5cents} to \eqref{eq:ineqdelta} then yields
%\begin{align}
%\beta_\alpha(\rho, \: \sigma) \geq \sum_{ij} s\min(t'\lambda_i, \mu_j)\bigl(\lvert \braket{x_i | \Pi_t | y_j}\rvert +  \notag \\  \lvert \braket{x_i | \bar{\Pi}_t | y_j}\rvert \bigr)^2- \frac{s}{1-s}t'\beta. \label{eq:ineqs}
%\end{align}
%Finally, using that for any $u_1, u_2 \in \mathbb{C}$, $\lvert u_1\rvert + \lvert u_2\rvert \geq \lvert u_1 + u_2\rvert$ and the fact that $\Pi_t + \bar{\Pi}_t = I$
\begin{align}
\beta_\alpha(\rho, \: \sigma) &\geq s\Biggl(\sum_{ij} \min(t'\lambda_i, \mu_j) \lvert\braket{x_i | y_j}\rvert^2  - \frac{1}{1-s}t'\alpha \biggr). \label{eq:ineqsclass}
\end{align}

We next relate the right-hand side of \eqref{eq:ineqsclass} to $\beta_{\alpha}\left(P, Q\right)$. Indeed, using the definitions of $P$ and $Q$ in \eqref{eq:Ns-mapping}, we note that
\begin{align}
 \min(t'\lambda_{i}, \mu_{j}) \lvert\braket{x_{i}|y_{j}}\rvert^2 &= t'P_{ij} \bm{1}_{\{t'\lambda_{i} <  \mu_{j}\}} 
 \!+ Q_{ij} \bm{1}_{\{t'\lambda_{i} \geq \mu_{j}\}},
 \label{eqn:ineqPQ}
\end{align}
where $\bm{1}_{\mathcal{E}}$ denotes the indicator function for the event $\mathcal{E}$. Specializing the variational formulation \eqref{eq:var} for $\rho, \sigma$ being diagonal operators with $P, Q$ in their diagonal yields %$\beta_{\alpha}\left(P, Q\right)$ can be written as
\begin{align}
 \beta_{\alpha}\left(P, Q\right) 
 &=  \sup_{t \geq 0} \Bigl\{ \sum_{ij} Q_{ij} \bm{1}_{\{tP_{ij} \geq Q_{ij}\}} \notag \\ &\qquad\qquad {} + t \Bigl( \sum_{ij} P_{ij} \bm{1}_{\{tP_{ij} < Q_{ij}\}} - \alpha \Bigr)\Bigr\}. \label{eq:this_too}
\end{align}
Combining \eqref{eqn:ineqPQ} with \eqref{eq:ineqsclass}, taking the supremum over $t' \geq 0$, and expressing the result using  \eqref{eq:this_too}, we obtain the bound \eqref{eq:thmlow}.
% \begin{align}
% \beta_{\alpha}(\rho, \sigma) \geq s \beta_{\frac{1}{1-s}\alpha}(P, Q).
% \end{align}
\end{IEEEproof}

\subsection{Pure-State Discrimination}
The bound in Theorem~\ref{thm:low} is asymptotically tight for a test between pure states that are asymptotically orthogonal. Consider the discrimination between the pure states
\begin{align}
 \mathrm{H}_0\colon \rho = \ket{x_1}\bra{x_1}, \qquad
 \mathrm{H}_1 \colon \sigma = \ket{y_1}\bra{y_1},\label{eq:HT-pure-H0-H1}
\end{align}
where $\ket{x_1}$ and $\ket{y_1}$ are assumed to satisfy $0<\left|\braket{x_1|y_1}\right|^2 < 1$.

\subsubsection{Classical test}
For the above quantum states $\rho$ and $\sigma$,
the Nussbaum-Szkoła mapping from \eqref{eq:Ns-mapping} becomes
\begin{align}
 P_{ij} = \begin{cases}
               \lvert\braket{x_1|y_j}\rvert^2 , \:\:\: i=1,\\
               0, \:\:\:\:\:\:\quad\text{otherwise},
           \end{cases} \,
 Q_{ij} = \begin{cases}
               \lvert\braket{x_i|y_1}\rvert^2 , \:\:\: j=1,\\
               0,\:\:\:\:\:\:\quad\text{otherwise}.
           \end{cases}
\end{align}
The distributions $P$ and $Q$ exhibit non-overlapping supports,
except in the singular case $(i,j)=(1,1)$, where 
\begin{align}
 P_{11} = Q_{11} = \lvert\braket{x_1|y_1}\rvert^2 = \Tr[\rho\sigma] \triangleq a. \label{eq:def_a}
\end{align}
%Here we defined $a = \lvert\braket{x_1|y_1}\rvert^2$ for future convenience.
The optimal classical test for this problem decides on the correct hypothesis with no error, except when $(i,j)=(1,1)$. In this case, the optimal test can select between $\mathrm{H}_0$ and $\mathrm{H}_1$ at random with some probabilities $p$ and $1-p$, thus incurring an error with probabilities
\begin{align}
  \alpha = (1-p) P_{11} = (1-p) a,\qquad
  \beta = p Q_{11} = p a.
\end{align}
After solving for $p$ in $\alpha$ and substituting into $\beta$, we obtain the error trade-off
\begin{align}
  \beta_{\alpha}(P,Q) = a - \alpha, \quad 0 \leq \alpha \leq a.\label{eq:trade-off-pure-c}
\end{align}

\subsubsection{Quantum test}
For the pure quantum states $\rho, \sigma$ given in \eqref{eq:HT-pure-H0-H1}, the quantum error trade-off is characterized in the following lemma.

\begin{lemma}\label{lem:pure-tradeoff}
Let $\rho, \sigma$ be pure states. Then, the optimal quantum error trade-off is given by
\begin{align}
\beta_{\alpha}(\rho, \sigma) = \alpha -2\alpha a + a - 2 \sqrt{(1-\alpha)\alpha(1-a)a}, \label{eq:pure_betaq}
\end{align}
for $0\leq \alpha \leq a$, where $a$ is given in \eqref{eq:def_a}.
%Moreover, for each $0 \leq p \leq 1$, the Lagrange multiplier associated with a specific $\alpha$ constraint is given by $t = p/(1-p)$.
\end{lemma}
\begin{IEEEproof}
The fidelity between two states is defined as
\begin{align}
    F(\rho, \sigma) \triangleq \Tr \left[ \lvert \sqrt{\rho}\sqrt{\sigma}\rvert\right], \label{eq:fidelity}
\end{align}
where $|A| = \sqrt{A^\dagger A}$.
Combining the fidelity lower bound from \cite[Eq.~(10)]{pereira2023analytical}, which is tight for pure states, with the fact that $F^2 = a$, the lemma follows.
\end{IEEEproof}

%We next write $\alpha$ as $\alpha= ca$ for some $0\leq c \leq 1$.
Expanding \eqref{eq:pure_betaq} as $a\to 0$, we obtain
\begin{align}
   % \beta_\alpha(\rho, \sigma) &= (c - 2\sqrt{c} + 1)a + o(a) \\
   \beta_\alpha(\rho, \sigma) = \alpha - 2\sqrt{\alpha a } + a + o(a), \quad 0 \leq \alpha \leq a,\label{eq:taylor}
\end{align}
where $o(a)$ summarizes terms that vanish faster than $a$ as \mbox{$a\to 0$}. Furthermore, using \eqref{eq:trade-off-pure-c}, the lower bound of Theorem~\ref{thm:low} can be written as
\begin{align}
    \beta_{\alpha}(\rho,\sigma) \geq s\beta_{\frac{\alpha}{1-s}}(P,Q) = sa - \frac{s}{1-s}\alpha, \quad 0\leq s < 1,
\end{align}
which, for $s=1-\sqrt{\frac{\alpha}{a}}$, becomes
\begin{align}
\beta_{\alpha}(\rho,\sigma) \geq \alpha - 2\sqrt{\alpha a} + a.\label{eq:pure-low-ab}
\end{align}
Comparing \eqref{eq:taylor} with \eqref{eq:pure-low-ab}, we observe that the quantum error trade-off in \eqref{eq:taylor} coincides with the lower bound in \eqref{eq:pure-low-ab} up to terms of order $o(a)$.
For a test between the $n$-th tensor product states $\rho^{\otimes n}, \: \sigma^{\otimes n}$, we have that $a= \lvert \braket{x_1 | y_1} \rvert^{2n}$. So the assumption $\lvert \braket{x_1 | y_1} \rvert^2 < 1$ implies that $a \to 0$ as $n \to \infty$. This proves that, for distinct pure states, the bound is asymptotically tight as $n\to\infty$. 

\begin{figure}
    \centering
    % Figure 1a
    \includegraphics[width=1\columnwidth]{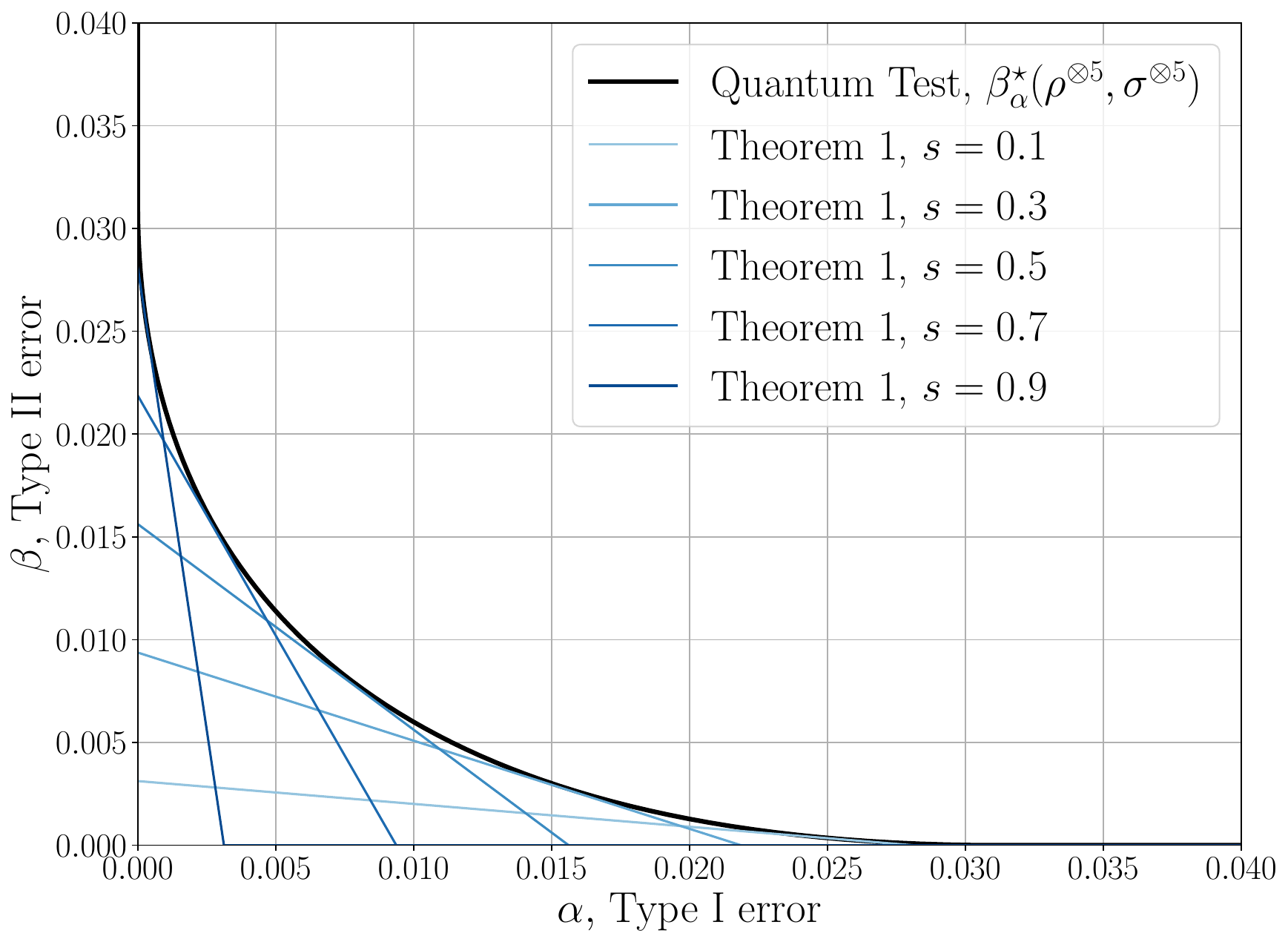}
    \caption{Quantum error trade-off compared to the lower bounds from Theorem~\ref{thm:low} for a hypothesis test between the states ${\ket{0}\!\bra{0}}^{\otimes 5}$ and ${\ket{+}\!\bra{+}}^{\otimes 5}$.}
    \label{fig:pure}
\end{figure}

As an example, we consider the $n$-th tensor product of the states  $\rho=\ket{0}\!\bra{0}$ and $\sigma = \ket{+}\!\bra{+}$. Fig.~\ref{fig:pure} shows the quantum error trade-off $\beta_{\alpha}$ for the hypothesis test between $\rho^{\otimes n}$ and $\sigma^{\otimes n}$ when $n=5$, and compares it to the lower bound of Theorem~\ref{thm:low} for different values of $s$. Observe that the upper envelope of the resulting bounds provides an accurate characterization of the quantum error trade-off when both states are pure and approximately orthogonal (in this example $a=2^{-5}\approx 0.03$). We therefore conclude that the  multiplicative factors $s$ and $1-s$ from Theorem~\ref{thm:low} cannot be improved in general.

%%%%%%%%%%%%%%%%%%%%%%%%%%%%%%%%%%%%%%%%%%%%%%%%%%%%%%%%%%%%%%%%%%%%%%%%%%%%%%%%%%%%%%%%%%%%%%%%%%%%%%%%
\section{Asymptotic Analysis}
%%%%%%%%%%%%%%%%%%%%%%%%%%%%%%%%%%%%%%%%%%%%%%%%%%%%%%%%%%%%%%%%%%%%%%%%%%%%%%%%%%%%%%%%%%%%%%%%%%%%%%%%
In this section, we show that the lower bound in Theorem~\ref{thm:low}, with an adequate choice of the parameter $s$, provides a unified and conceptually simple framework for recovering converse results in the asymptotic small-deviations (second-order), moderate-deviations, and large-deviations regimes. To this end, we extend the hypothesis test from single copies $\rho, \sigma$ to $n$-fold tensor products $\rho^{\otimes n}, \sigma^{\otimes n}$.
Furthermore, we introduce two relevant classical quantities associated with the Nussbaum-Szkoła-mapping distributions $P$ and $Q$, namely,
\begin{align}
V(P\|Q) & \triangleq \Exp_P\Biggl[\biggl(\log\frac{P}{Q}-D(P\|Q)\biggr)^2\Biggr] \\
T(P\|Q) & \triangleq \Exp_P\Biggl[\biggl|\log\frac{P}{Q}-D(P\|Q)\biggr|^3\Biggr]
\end{align}
where $D(P\|Q)$ denotes the relative entropy between $P$ and~$Q$. We further define the hypothesis testing relative entropy
\begin{align}
    D_h^{\varepsilon}(\rho^{\otimes n} \| \sigma^{\otimes n}) &\triangleq 
    -\log\beta_{\varepsilon}\bigl(\rho^{\otimes n}, \sigma^{\otimes n}\bigr). \label{eq:hrel}
\end{align}
For the classical Nussbaum-Szkoła distributions $P$ and $Q$, $D_h^{\varepsilon}(P^n \| Q^n)$ is defined analogously.

%%%%%%%%%%%%%%%%%%%%%%%%%%%%%%%%%%%%%%%%%%%%%%%%%%%%%%%%%%%%%%%%%%%%%%
\begin{corollary}[Small Deviations]\label{thm:small-deviations}
    Let $V(P \: \| \: Q) > 0$. Then,
    \begin{align}
        D_h^{\varepsilon}(\rho^{\otimes n} \: \| \: \sigma^{\otimes n}) &\leq  nD(P \: \| \: Q) + \sqrt{n V(P \: \| \: Q)} \Phi^{-1}\left( \varepsilon \right) \notag \\ &\quad + O(\log{n}), \label{eq:prop1}
    \end{align}
    where $\Phi(x)$ denotes the cumulative distribution function of the standard normal distribution and $O(\log n)$ summarizes terms of order $\log n$.
\end{corollary}
\begin{IEEEproof}%[Proof of Theorem~\ref{thm:small}] 
Taking logarithms on both sides of \eqref{eq:thmlow}, and replacing $s$ by $s_n$ to indicate that this parameter may depend on $n$, we obtain from Theorem~\ref{thm:low} that
\begin{align}
    D_h^{\varepsilon}(\rho^{\otimes n} \: \| \: \sigma^{\otimes n}) \leq D_h^{\frac{\varepsilon}{1-s_n}}(P^n \: \| \: Q^n) -\log{s_n}.\label{eq:thm1rel}
\end{align}
From the small-deviations asymptotic behavior of the classical hypothesis testing problem \cite[Prop.~2.3]{tan2014asymptotic}, we have
\begin{align}
    D_h^{\varepsilon}(P^n \: \| \: Q^n) %\notag\\
    &= nD(P \: \| \: Q) + \sqrt{n V(P \: \| \: Q)} \Phi^{-1}\left( \varepsilon \right)
    \notag \\ & \quad
    {} + \frac{1}{2}\log{n}  + O(1). \label{eq:prop23}
\end{align}
Using \eqref{eq:prop23} in \eqref{eq:thm1rel}, and setting $s_n = 1/{\sqrt{n}}$, we obtain that
\begin{align}
    &D_h^{\varepsilon}(\rho^{\otimes n} \| \sigma^{\otimes n}) \leq nD(P \| Q) 
    \notag \\
    &{} +\sqrt{n V(P \: \| \: Q)} \Phi^{-1}\left( \frac{\varepsilon}{1-n^{-1/2}} \right)
    + \log{n} + O(1). \label{eq:relentuppsmall1}
\end{align}
Performing a first-order Taylor expansion of $\Phi^{-1}( \frac{\varepsilon}{1-n^{-1/2}})$ around $\varepsilon$, and grouping terms of order $\log n$ and $1$ in one $O(\log n)$-term, proves the corollary.
%\begin{align}
%    \Phi^{-1}\left( \frac{\varepsilon}{1-n^{-1/2}} \right) 
%    =  \Phi^{-1}( \varepsilon ) + O(n^{-1/2}),
%\end{align}
%and note that $n^{1/2} O(n^{-1/2}) = O(1)$.
%Then, grouping the terms $\frac{1}{2}\log{n} + \frac{1}{2}\log{n} + O(1) = O(\log{n})$
%completes the proof.
\end{IEEEproof}

%%%%%%%%%%%%%%%%%%%%%%%%%%%%%%%%%%%%%%%%%%%%%%%%%%%%%%%%%%%%%%%%%%%%%
%\subsection{Moderate Deviation Analysis} 

\begin{corollary}[Moderate Deviations]\label{thm:moderate-deviations}
Let $\{a_n\}$ be a nonnegative sequence satisfying $a_n \to 0$ and
$n a_n^2 \to\infty$ as $n\to\infty$. Further let $\varepsilon_n = e^{-n a_n^2}$. Then,
\begin{equation} \label{eq:moderate}
D_h^{\varepsilon_n}(\rho^{\otimes n} \| \sigma^{\otimes n}) \leq nD(P\|Q) - \sqrt{2 V(P\|Q)} n a_n + o(n a_n)
\end{equation}
where $o(na_n)$ summarizes terms that grow more slowly than $na_n$ as $n\to\infty$.
\end{corollary}
\begin{IEEEproof}
We start by upper-bounding $D_h^{\varepsilon_n}(\rho^{\otimes n} \| \sigma^{\otimes n})$
using Theorem~\ref{thm:low} with $s \equiv s_n$ and applying Lemma~1 to the term $\beta_{\frac{1}{1-s_n}\varepsilon_n}(P^n, Q^n)$.
We then set $t= 2^{-R_n}$ and use that $Q^n\bigl(\log\frac{P^n}{Q^n} > R_n\bigr) \geq 0$ to obtain the upper bound
\begin{align}
&D_h^{\varepsilon_n}(\rho^{\otimes n} \| \sigma^{\otimes n})
\notag\\&
 \leq R_n - \log\left(P^n\!\left(\!\log\frac{P^n}{Q^n} \leq R_n\!\right)
  - \frac{1}{1-s_n}\varepsilon_n \right) - \log s_n \label{eq:UB}
\end{align}
for arbitrary sequences $\{R_n\}$ and $\{s_n\}$. %To prove the moderate-deviations result, 
Specifically, we set
\begin{align}
R_n = n D(P\|Q) - n a_n \lambda_n, \quad 
s_n = \frac{1}{n a_n}, \label{eq:sn_Rn}
\end{align}
for a sequence $\{\lambda_n\}$ that satisfies
\begin{align}
\lambda_n^2 &= 2 V(P\|Q) + o(1) \label{eq:lambda_asymp}
\end{align}
and
\begin{align}
P^n\biggl(\log\frac{P^n}{Q^n} \leq R_n\biggr) &\geq \frac{1}{1-2s_n} \varepsilon_n. \label{eq:LB_P}
\end{align}
Using \eqref{eq:LB_P} and $\varepsilon_n = e^{-n a_n^2}$, and noting that $n a_n = \sqrt{n}\sqrt{na_n^2}\to\infty$ as $n\to\infty$ by the assumption $na_n^2\to\infty$, the last two terms in \eqref{eq:UB} become
\begin{align}
 &-\log\biggl(P^n\biggl(\log\frac{P^n}{Q^n} \leq R_n\biggr)-\frac{1}{1-s_n}\varepsilon_n\biggr) -\log s_n 
 \notag \\ &\quad\;\;\qquad\qquad\qquad\qquad\quad\qquad\quad\qquad\qquad 
 = o(n a_n). \label{eq:forget_logP}
\end{align}
Substituting \eqref{eq:sn_Rn}, \eqref{eq:lambda_asymp}, and \eqref{eq:forget_logP} in \eqref{eq:UB}, 
and noting that $\sqrt{2V(P\|Q)+o(1)} = \sqrt{2 V(P\|Q)} + o(1)$,
we obtain the desired bound \eqref{eq:moderate}.

It remains to show that there exists a sequence $\{\lambda_n\}$ that satisfies \eqref{eq:lambda_asymp}
and \eqref{eq:LB_P}. Specifically, we choose
\begin{align}
\lambda_n^2 = 2 V(P\|Q)\left(1-\frac{A_n}{n a_n^2} - \frac{B_n}{n a_n^2}+ \frac{\log(1-2 s_n)}{n a_n^2}\right) \label{eq:def_lambda}
\end{align}
where
\begin{align}
A_n &\triangleq \frac{c_1\sqrt{8}T(P\|Q)}{V^{3/2}(P\|Q)}n a_n^3 - \ln\biggl(\!1-\frac{c_2\sqrt{2} T(P\|Q)}{V(P\|Q)}a_n\!\biggr) \\
B_n &\triangleq \frac{1}{2}\ln(2n a_n^2) - \ln\biggl(\!1-\frac{1}{n a_n^2}\!\biggr)
\end{align}
with constants $c_1, c_2>0$. Note that the sequences $\{A_n\}$ and $\{B_n\}$ are nonnegative and of order $o(n a_n^2)$ by the assumption $a_n\to 0$. Moreover, for our choice of $s_n=1/(n a_n)$, we have that $\ln(1-2 s_n) = o(1)$, since $n a_n\to\infty$ as $n\to\infty$. Consequently, this choice of $\{\lambda_n\}$ satisfies \eqref{eq:lambda_asymp}.

To show that $\{\lambda_n\}$ also satisfies \eqref{eq:LB_P}, we apply the following result due to Rozovsky:
\begin{lemma}[\hspace{-0.01mm}{\cite[Th.~B2]{rozovsky2002estimate}}]\label{lem:rozovsky}
 Let $X_1,\ldots,X_n$ be independent random variables with finite third moments
\begin{align}
\!\mu_k = \E{X_k}, \;\, \sigma_k^2 = \Var{X_k}, \;\, t_k = \E{|X_k-\mu_k|^3}.
\end{align}
Let $S= \sigma_1^2+\ldots+\sigma_n^2$ and $T=t_1+\ldots+t_n$, and let $c_1>0$ and $c_2>0$ be universal constants. Then, for every $x \geq 1$, 
\begin{align}
\!&\Prob\Biggl(\sum_{k=1}^n (X_k - \mu_k) > x \sqrt{S}\Biggr) 
%\notag \\ & \qquad \qquad \qquad \quad \quad 
\geq Q(x) e^{-\frac{c_1 T}{S^{3/2}}x^3}\biggl(1-\frac{c_2 T}{S^{3/2}}x\biggr). \label{eq:rozovsky}
\end{align}
\end{lemma}

Specifically, we consider
\begin{align}
X_k & = -\log\frac{P}{Q}, & \mu_k & = - D(P\|Q), \notag \\ 
\sigma_k^2 & = V(P\|Q), & t_k & = T(P\|Q), \label{eq:def_roz}
\end{align}
so that 
the probability on the left-hand side of \eqref{eq:LB_P} becomes
\begin{align}
\!P^n\biggl(\log\frac{P^n}{Q^n} \leq R_n\biggr) =  \Prob\Biggl(\sum_{k=1}^n (X_k-\mu_k) \geq \lambda_n n a_n\Biggr).\label{eq:prob_roz}
\end{align}
We apply Lemma~\ref{lem:rozovsky} with $S = n V(P\|Q)$ and $T=n T(P\|Q)$ to~\eqref{eq:prob_roz} and use that, by the definition of $\lambda_n^2$ in \eqref{eq:def_lambda}, $V(P\|Q) \leq \lambda_n^2 \leq 2 V(P\|Q)$.
Taking natural logarithms on both sides of \eqref{eq:prob_roz}, and further using the lower bound $Q(x) \geq 1/\sqrt{2\pi x^2} e^{-x^2/2}(1-x^{-2})$, $x>0$, we obtain that
\begin{align}
\ln P^n\biggl(\log\frac{P^n}{Q^n} \leq R_n\biggr) 
&\geq -n a_n^2 - \ln(1-2 s_n). \label{eq:LB_this}
\end{align}
Therefore,
\begin{align}
P^n\biggl(\log\frac{P^n}{Q^n} \leq R_n\biggr) 
&\geq e^{-n a_n^2-\ln(1-2 s_n)}
\notag\\
&= \frac{1}{1-2 s_n} \varepsilon_n,
\end{align}
which is \eqref{eq:LB_P}.
\end{IEEEproof}

%%%%%%%%%%%%%%%%%%%%%%%%%%%%%%%%%%%%%%%%%%%%%%%%%%%%%%%%%%%%%%%%%%%%%%%
\begin{corollary}[Large Deviations]\label{thm:large}
    Let $\varepsilon_n \leq e^{-n r}$. Then,
\begin{align}
  &\mathop{\lim\sup}_{n\to\infty} \frac{1}{n} D_h^{\varepsilon_n}(\rho^{\otimes n} \| \sigma^{\otimes n})\notag\\
    &\qquad\;\leq  \sup_{0\leq s\leq 1} \left\{ \frac{1}{s-1} \log \Tr\bigl[ \rho^s \sigma^{1-s}\bigr] + \frac{s}{s-1} r \right\}.\label{eq:hoeffding-bound}
\end{align}
\end{corollary}
\begin{IEEEproof}
Setting $s = \frac{1}{2}$ in Theorem~\ref{thm:low} recovers the lower bound in \cite[Prop.~2]{audenaert2008} (for details, see~\cite{vazquez2024error}). The quantum Hoeffding bound~\eqref{eq:hoeffding-bound} is then obtained by following the steps in the proof of \cite[Th.~3]{audenaert2008}.
\end{IEEEproof}

\section{Comparison with Previous Converse Bounds}
In this section, we compare the converse bound of Theorem~\ref{thm:low} with previous converse bounds obtained in the literature.
In particular, we consider the information-spectrum bound in \cite[Eq.~(27)]{tomamichel2013hierarchy}, the fidelity bound in \cite[Eq.~(10)]{pereira2023analytical}, and the Nussbaum-Szkoła bound on the minimum average error probability \cite[Prop.~2]{audenaert2008} which, extended to the asymmetric setting, coincides with Theorem~\ref{thm:low} with $s=1/2$.

\begin{figure}
    \centering
    % Figure 1a
    \includegraphics[width=1\columnwidth]{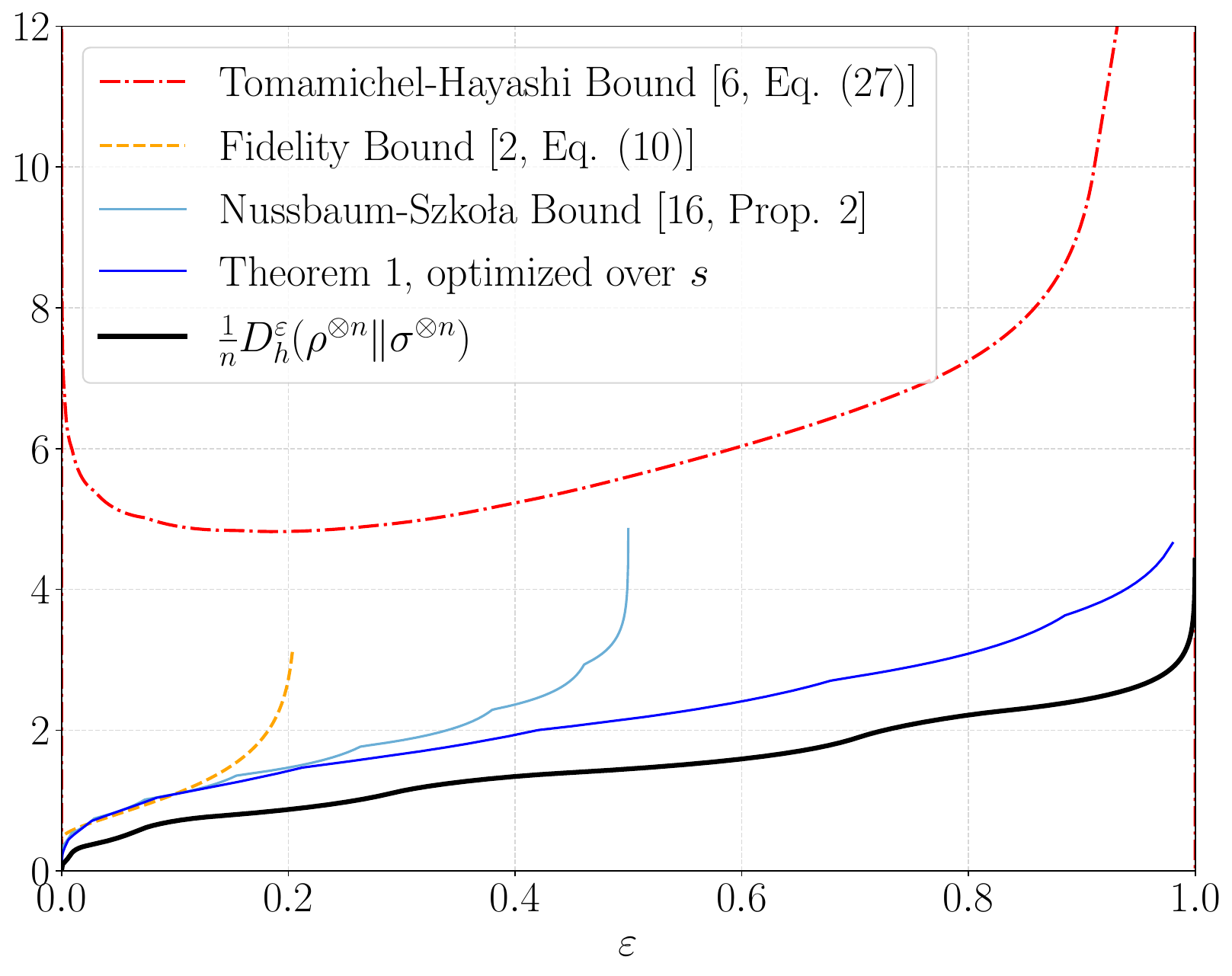}
    \caption{Upper bounds on the normalized hypothesis testing relative entropy $\frac{1}{n}D_h^{\varepsilon}(\rho^{\otimes n} \| \sigma^{\otimes n})$ for $\rho,\sigma$ in \eqref{eq:mixedstates} and $n=5$.} 
    \label{fig:HTrelative_n5}
\end{figure}

Figure~\ref{fig:HTrelative_n5} compares the resulting upper bounds with the true normalized hypothesis testing relative entropy
\begin{align}
    \frac{1}{n}D_h^{\varepsilon}(\rho^{\otimes n} \| \sigma^{\otimes n}) &=
    -\frac{1}{n}\log\beta_{\varepsilon}\bigl(\rho^{\otimes n}, \sigma^{\otimes n}\bigr)
\end{align}
for $n=5$ and the mixed states
\begin{align}
\rho = \begin{bmatrix} 0.8 & 0 \\ 0 & 0.2 \end{bmatrix}, \qquad \sigma =\begin{bmatrix} 0.35 & \frac{3\sqrt{3}}{20} \\ \frac{3\sqrt{3}}{20} & 0.65\end{bmatrix}. \label{eq:mixedstates}
\end{align}
Observe the limited range of the bounds in \cite[Prop.~2]{audenaert2008} and \cite[Eq.~(10)]{pereira2023analytical},
which yield finite non-trivial bounds only for $0 \leq \varepsilon < F(\rho, \sigma)^{2n}$ and $0 \leq \varepsilon < 1/2$, respectively.
In contrast, the bound of Theorem~\ref{thm:low}, optimized over $s\in[0,1]$, 
provides a tight characterization of $\frac{1}{n}D_h^{\varepsilon}(\rho^{\otimes n} \| \sigma^{\otimes n})$  over the entire range $0 \leq \varepsilon < 1$. Furthermore, the bound \cite[Eq.~(27)]{tomamichel2013hierarchy} yields finite values over $0 < \varepsilon < 1$, but it is significantly looser than the bound obtained from Theorem~\ref{thm:low} for the considered example.

We further note that, while the bound in \cite[Eq.~(27)]{tomamichel2013hierarchy} recovers the small- and moderate-deviation results of Corollaries~\ref{thm:small-deviations} and \ref{thm:moderate-deviations}
(see~\cite{tomamichel2013hierarchy} and \cite{chubb2017moderate}),
it fails to recover the error exponent in the large-deviations analysis beyond Stein's regime.
We conclude that the bound presented in Theorem~\ref{thm:low} not only constitutes an accurate non-asymptotic approximation of the quantum hypothesis testing performance, it also provides a unified framework for recovering converse results in the asymptotic setting.

\pagebreak

\balance

\bibliographystyle{IEEEtran}
\bibliography{bib/references}

@article{audenaert2007discriminating,
  title={Discriminating States: The Quantum {Chernoff} Bound},
  author={Audenaert, Koen M. R. and Calsamiglia, John and Munoz-Tapia, Ramon and Bagan, Emilio and Masanes, Lluis and Acin, Antonio and Verstraete, Frank},
  journal={Physical Review Letters},
  volume={98},
  number={16},
  pages={160501},
  year={2007},
  month={Apr.},
  publisher={American Physical Society}
}

@article{audenaert2008,
  title = {Asymptotic Error Rates in Quantum Hypothesis Testing},
  author = {Audenaert, K. M. R. and Nussbaum, M. and Szkoła, A. and Verstraete, F.},
  journal = {Communications in Mathematical Physics},
  volume = {279},
  number = {1},
  pages = {251--283},
  year = {2008},
  month = {Apr.},
  day = {01},
  issn = {1432-0916},
}

@article{hiaipetz1991,
  title={The proper formula for relative entropy and its asymptotics in quantum probability},
  author={Hiai, Fumio and Petz, Dénes},
  journal={Communications in Mathematical Physics},
  volume={143},
  number={1},
  pages={99--114},
  year={1991},
  month ={Dec.},
  publisher={Springer}
}

@article{ogawahayashi2004,
  title={On error exponents in quantum hypothesis testing},
  author={Ogawa, Tomohiro and Hayashi, Masahito},
  journal={IEEE Transactions on Information Theory},
  volume={50},
  number={6},
  pages={1368--1372},
  year={2004},
    month={Jun.},
  publisher={IEEE}
}

@article{nagaoka2006converse,
  title={The Converse Part of The Theorem for Quantum {H}oeffding Bound},
  author={Hiroshi Nagaoka},
  journal={arXiv: Quantum Physics},
  year={2006},
}

@article{hayashi2006error,
  title = {Error exponent in asymmetric quantum hypothesis testing and its application to classical-quantum channel coding},
  author = {Hayashi, Masahito},
  journal = {Physical Review A},
  volume = {76},
  issue = {6},
  pages = {062301},
  numpages = {4},
  year = {2007},
  month = {Dec},
  publisher = {American Physical Society},
  doi = {10.1103/PhysRevA.76.062301},
}

@article{pereira2023analytical,
  title={Analytical bounds for nonasymptotic asymmetric state discrimination},
  author={Pereira, Jason L and Banchi, Leonardo and Pirandola, Stefano},
  journal={Physical Review Applied},
  volume={19},
  number={5},
  pages={054030},
  year={2023},
  month = {May},
  publisher={APS}
}

@article{tomamichel2013hierarchy,
  title={A Hierarchy of Information Quantities for Finite Blocklength Analysis},
  author={Tomamichel, Marco and Hayashi, Masahito},
  journal={IEEE Transactions on Information Theory},
  volume={59},
  number={10},
  pages={6719--6736},
  year={2013},
  month={Nov.},
  publisher={IEEE}
}

@article{tan2014asymptotic,
  title={Asymptotic estimates in information theory with non-vanishing error probabilities},
  author={Tan, Vincent YF},
  journal={Foundations and Trends{\textregistered} in Communications and Information Theory},
  volume={11},
  number={1-2},
  pages={1--184},
  year={2014},
  publisher={Now Publishers, Inc.}
}

@article{chubb2017moderate,
  title={Moderate deviation analysis for classical communication over quantum channels},
  author={Chubb, Christopher T and Tan, Vincent YF and Tomamichel, Marco},
  journal={Communications in Mathematical Physics},
  volume={355},
  number={3},
  pages={1283--1315},
  year={2017},
  month={Aug.},
  publisher={Springer}
}

@article{nussbaum2009lower,
    author = {Michael Nussbaum and Arleta Szkoła},
    title = {{The Chernoff lower bound for symmetric quantum hypothesis testing}},
    volume = {37},
    journal = {The Annals of Statistics},
    number = {2},
    publisher = {Institute of Mathematical Statistics},
    pages = {1040 -- 1057},
    year = {2009},
    month = {Apr.},
    doi = {10.1214/08-AOS593}
}

@article{rozovsky2002estimate,
  title={Estimate from below for large-deviation probabilities of a sum of independent random variables with finite variances},
  author={Rozovsky, LV},
  journal={Journal of Mathematical Sciences},
  volume={109},
  number={6},
  pages={2192--2209},
  year={2002},
  month={May},
  publisher={Springer}
}

@article{cheng2017moderate,
  title={Moderate deviation analysis for classical-quantum channels and quantum hypothesis testing},
  author={Cheng, Hao-Chung and Hsieh, Min-Hsiu},
  journal={IEEE Transactions on Information Theory},
  volume={64},
  number={2},
  pages={1385--1403},
  year={2017},
  month={Feb.},
  publisher={IEEE}
}

@inproceedings{sason2012moderate,
  title={Moderate deviations analysis of binary hypothesis testing},
  author={Sason, Igal},
  booktitle={Proceedings 2012 IEEE International Symposium on Information Theory},
  pages={821--825},
  year={2012},
  month={Jul.},
  address={Boston, MA, USA}
}

@inproceedings{vazquez2024error,
  title={Error Probability Trade-off in Quantum Hypothesis Testing via the {N}ussbaum--{S}zko{\l}a Mapping},
  author={Vazquez-Vilar, Gonzalo and Lizarribar-Carrillo, Jorge},
  booktitle={Proceedings International Zurich Seminar on Information and Communication},
  pages={58--61},
  year={2024},
  month={Mar.},
  address={Zurich, Switzerland}
}

@article{li2014second,
  title={Second-order asymptotics for quantum hypothesis testing},
  journal={The Annals of Statistics},
  volume={42},
  number={1},
  pages={171--189},
  author={Li, Ke},
  year={2014},
  month={Feb.}
}

@inproceedings{vazquez2016multiple,
  title={Multiple quantum hypothesis testing expressions and classical-quantum channel converse bounds},
  author={Vazquez-Vilar, Gonzalo},
  booktitle={Proceedings 2016 IEEE International Symposium on Information Theory},
  pages={2854--2857},
  year={2016},
  month={Jul.},
  address={Barcelona, Spain}
}
\end{document}